\documentclass[12pt]{article}

\def\fun#1#2{\lower3.6pt\vbox{\baselineskip0pt\lineskip.9pt
\ialign{$\mathsurround=0pt#1\hfil##\hfil$\crcr#2\crcr\sim\crcr}}}

\title{Remarks on Higgs boson production}
\author{A.Dobrovolskaya, V.Novikov\\
ITEP 117218 Moscow, Russia}
\date{}
\begin{document}
\maketitle
\vskip 4cm
\begin{abstract}

We demonstrate that Higgs boson production in fusion processes is
accompanied by large azimuthal asymmetry. This asymmetry can be used 
to facilitate detection of Higgs boson events. We also demonstrate 
that the double Higgs boson production at threshold is a very appropriate
process to test the existence of new physics at higher energy.

\end{abstract}

\newpage

\section{Introduction.}

  In the modern theory of electroweak interaction Higgs boson plays
the very important role at high energies. Without Higgs bosons the
scattering amplitudes of massive $W$ and $Z$ bosons grow with energy
and
violate unitarity. Diagrams with Higgs bosons exchange improve
drastically this high-energy behaviour and protect  amplitudes  from
violation of unitarity \cite{1}.

 In contrast to the high-energy phenomena the contribution of Higgs
bosons into low-energy processes is very small according to the
Veltman's screening theorem \cite{2}. As a result a little is known
about Higgs
boson properties from the current experiments. From all electroweak
measurements at LEP1,SLC and LEP2 it follows that \cite{3}
$$
70 \mbox{\rm Gev} < m_h < 465 \mbox{\rm Gev} \;\;\; 95\% c.l.
$$
and that is it.

 The main mission of the future experiments is to discover Higgs boson
and to study its interactions.

 There are two different classes of such interactions. The first one is the
class of Higgs couplings with quarks, leptons, $W$ and $Z$ bosons.
These couplings are fixed by renormalizability of the theory. Any
change in Yukawa interaction
of Higgs boson with top quark or with $W$ and $Z$ bosons would
immediately
destroy the agreement of the loop calculation with precision electroweak data
even if one introduce some reasonable effective cut-off. Any new physics
can't produce sizable change of these effective couplings from canonical ones.

 The second class includes higgs self-interactions. In the Standard Model
$3H$ and $4H$ vertices are determined by the value of higgs
condensate and by
the value of higgs mass. Though renormalizability works in this case as well,
the new interactions can change 3- and 4-effective coupling constants from
the canonical values and this change will produce only tiny effect for low
energy physics. So phenomenologically  Higgs self-interaction coupling
constants are independent parameters. Experimental study of these parameters
would provide information about new physics at higher scales.

 In this talk I would like to present a few remarks concerning Higgs
boson production and the ways to study  effective Higgs boson
self-interaction.

\section{One Higgs boson production.}

 For the moderate value of Higgs mass ($m_h < 200$ Gev) the main
mechanism for Higgs boson production in colliders is Higgs-strahlung
process  $e^+ e^- \to (Z^*) \to Z H$. The corresponding
cross-sections drops as $1/s$ , where $s$ is  the c.m. energy of
beams.

 For higher higgs mass  the main mechanism for $H$-production is the
 fusion
mechanism. In this case  at first stage the effective $W$, $Z$ (or
gluon) beams are
produced by colliding particles  and then at the second stage Higgs
bosons are produced in the collisions of $WW$, $ZZ$  or gluon-gluon
effective beams.
The corresponding cross sections have very weak dependence on energy
$$
\sigma \sim const \cdot \log (s/m_h^2) \;\; .
$$

For example in $e^+ e^-$ collision  the $WW$-fusion cross section of
$H$-production
varies from 10 to 100 fb for all realistic values of  $s/m_h^2$. This
cross
section is two or three order of magnitude smaller than $e^+ e^-$
annihilation
cross section into hadrons. So it is rather difficult problem to extract
the events with Higgs boson production from the vast background events.

  Our first remark is that the best way to look for higgs events
is not to measure the total cross section but the special $\phi$
dependent terms in the differential cross sections. Here $\phi$
is the azimuthal angle
between scattering planes of initial scattered fermions.

 Indeed in the case of fusion production of Higgs bosons the cross
 section has a very strong  azimuthal asymmetry \cite{4}
\begin{equation}
\sigma_H =\sigma_1 (1+A_1\cos\phi +A_2 \cos 2\phi)
\end{equation}

It is important that numerically the factor $A_1$ is very large
in the whole
region of expected Higgs mass and beam energies $s$ \cite{4,5}
$$
A_1 \sim  0.5 - 0.7 \;\; .
$$
(factor $A_2$ is much smaller; $A_2 \sim 10^{-2}$).

 Parametrically $A_1$ and $A_2$  depend on $m_h$ and energy  $s$ :
 $A=A(m_h, s)$.
For heavy higgs and large energy $s\gg m_h^2$ one can find \cite{4}
$$
A_1 = \pi^2 (m_W^2/m_H^2) \;\; ,
$$
$$
A_2 \sim (m_W^2/m_H^2)^2 \;\; .
$$
So in the case of $H$-production the first and the second terms
(harmonics) in eq.(1) are of the same order of magnitude.

For the background events the  cross section is also a sum of three
harmonics
\begin{equation}
\sigma_{Bkgd} =\sigma_2(1+B_1 \cos\phi + B_2 \cos 2\phi)
\end{equation}
but the value of the coefficients in eq.(2) is quite different.
The first term in eq. (2) - the total cross section $\sigma_2$ - is
two or three order of
magnitude larger than $\sigma_1$ in eq.(1). The third
term is also large; as a rule
$B_2 \sim 1$. As for the second term in eq.(2) it is extremely small.
Parametrically factor $B_1$  is of the order of
$$
B_1 \sim <q^2>/<s> \;\; ,
$$
where  $<s>$ is the mean value of energy  of the effective $W$
beams and $<q^2>$ is
the mean value of virtual momenta of W. Thus for  production of light hadron
$$
B_1 \sim \Lambda_{QCD}^2/<s> \leq 10^{-4} \;\; ,
$$
i.e. it is small correction to the corresponding $\phi$-dependent
term in $H$-production cross section. For lepton $L$ and heavy quark
$Q$ production
$$
B_1 \sim m_{L,Q}^2 /<s>
$$
and also is a small correction to the $H$-production for all cases
with one exception - top-quark production has to be considered separately.

 So our statement is that the best way to look for $H$-production
events is to measure  the first harmonic, i.e. $\cos\phi$  term in
the differential cross section but not the total cross section.
It seems that experimentally it is quite possible to separate
one harmonics  from another one.

 The physical reason for such behaviour is rather simple. The
 $\phi$ -dependent terms in eqs.(1)-(2) originate from the interference
 of the amplitudes with different helicities of $W(Z)$ bosons. The
 considered $\cos\phi$ term corresponds
to spin-flip interference from helicity 1 to helicity 0 states.
The polarization vector $\varepsilon_{\mu}(0)$  of $W$, $Z$ or gluon
corresponding to the state with
helicity 0 is proportional to the momenta $q$
\begin{equation}
\varepsilon_{\mu}(0) =q_{\mu}/\sqrt{-q^2} +O (\frac{\sqrt{-q^2}}{s})
\end{equation}

 Vector bosons interact with conserved or partially conserved currents
of fermions, as a result the large component $\sim q_{\mu}$ in
eq.(3) is washed out
from the amplitudes of quarks and leptons production. As for the
$HWW$ or $HZZ$
verteces they are not transversal and  large components $\sim
q_{\mu}$ gives
100\% contribution into $H$-production.

  We repeat in conclusion that if one can separate experimentally three
harmonics from each other he will find that the first harmonic is enriched
enormously by Higgs boson events.

\section{Experimental study of Higgs selfinteraction.}

Self-interaction coupling constants can be measured only in
the multy Higgs bosons processes. The first nontrivial
sample is double Higgs boson production.

Corresponding cross section is very small in SM
$$
\sigma(2H) \sim 1 fb \;\;\; \mbox{\rm for} \;\;\; \sqrt{s} \sim 1 \;
\mbox {\rm TeV} \;\;, \;\; m_H \sim 100 \; \mbox{\rm GeV} \;\; .
$$
Nevertheless sooner or later we have to start to study such
processes.

Our second remark is that "new physics" beyond the SM can enhance
enormously the double Higgs boson production at threshold.

Indeed in the SM we have a very special choice of parameters.
Potential for Higgs boson doublet $\varphi$ depends only on two
parameters
\begin{equation}
V_{SM} =\frac{\lambda}{4}(|\varphi\varphi^+|^2 -\eta^2)^2
\end{equation}
On the other hand $V$ describes four physical quantities:
\begin{eqnarray}
\mbox{\rm higgs~~~ condensate:}& \eta \;\; ;  \nonumber    \\
\mbox{\rm higgs~~~ mass:}& m_H =\lambda\eta \;\; ;  \nonumber    \\
\mbox{\rm 3H~~~ vertex:}& \lambda_3(\frac{m_H^2}{2\eta})H^3 \;\; ;
\\
\mbox{\rm 4H~~~ vertex:}& \lambda_4(\frac{m_H^2}{8\eta^2})H^4 \;\; .
\nonumber
\end{eqnarray}

In the case of Standard Model
$$
\lambda_3 = \lambda_4 \equiv 1
$$

For general effective potential $V_{eff}(\varphi)$ these four
parameters are independent and factors $\lambda_3$ and $\lambda_4$
are out of tune
\begin{equation}
\lambda_3 \neq \lambda_4 \neq 1 \;\; .
\end{equation}

So one can expect some special relations for Higgs boson amplitudes
that take place in SM and that are invalid beyond the SM. Indeed this
is the case. The simplest example is double Higgs boson production in
the beams of longitudinal $W$ bosons. Corresponding amplitude
vanishes at threshold in the SM in tree approximation \cite{4}.
Moreover the amplitude with $n$ Higgs boson in final state goes to
zero at threshold as well \cite{6}.

This fine-tuning cancellation of different diagrams at threshold
takes place only if $\lambda_3 =\lambda_4 =1$. If new physics at
scale $\Lambda \gg m_h$ modifies Higgs boson effective potential at
low energy the exact cancellation will be destroyed. So, in general,
any new physics works only in favorable direction -- it enhances
cross section for Higgs production at threshold.

Of course, it is possible that "new physics" is screened at low
energy by small factor $(m_h^2/\Lambda^2)$ \cite{7}. In this case the
influence of "point-like" "new physics" at threshold effects is small
and we have no chance to learn a lot from the experiments at threshold.

It is also possible that large scale $\Lambda$ has "soft", infrared
origin \cite{8}. In this case if new interaction can be treated
perturbatively the effective low-energy potential for Higgs field
$H(x)$ has a form
\begin{equation}
\delta V_{eff} = \frac{1}{4}\gamma H^4(x) \ln
\frac{H^2(x)}{\Lambda^2}
\end{equation}
It is a general form for any perturbative one-loop effective
potential, where factor $\gamma$ is connected with $\beta$-function
of new coupling constants and $\Lambda$ determines the scale of new
interaction. For this potential the effective 3 and 4-vertex factors
$\lambda_{3,4}$ are rather arbitrary, but not absolutely arbitrary.

Stability of the Standard Model vacuum that corresponds to
$$
<H> = 246 \; \mbox{\rm GeV}
$$
takes place only if effective 3-vertex $\lambda_{3}$ varies in the
region
$$
1 < \lambda_3 < 7/3 \;\; .
$$

For any value $\lambda_3 \neq 1$ the $2H$ boson amplitude does not
vanish at threshold. Since the energy of effective $W$ bosons is not
fixed but it has some distribution we can't see this infinite threshold
enhancement. It is replaced by finite, but large factor. For $\lambda_3
= 7/3$ the estimates show that this factor is of the order of 5 to
10 \cite{8}.

\section{Conclusion.}

We have analyzed the $\phi$ dependence of single Higgs boson
production cross section and demonstrated that the first harmonic
proportional to $\cos\phi$ is enriched by Higgs boson events.

We also demonstrated that the double Higgs boson production at
threshold is the very appropriate process to test the existence of
new non-standard physics at higher scales.

\section{Acknowledgments.}

This work is partially supported by INTAS-RFBR grant 95-0567.


\begin{thebibliography}{99}
\bibitem{1} B.Lee, C.Quigg and H.Thacker, Phys. Rev. {\bf D16}, 1514
(1977).
\bibitem{2} M.Veltman, Nucl. Phys. {\bf B213}, 89 (1977).
\bibitem{3} A.B\"{o}hm, Result from the Measurements of Electroweak
Processes at LEP1, talk at XXIInd Rencontre de Moriond, Les Arcs 1,
Savoie, France, March 15-22, 1997.
\bibitem{4} A.Dobrovolskaya, V.Novikov, Z.Phys. {\bf C52}, 427
(1991).
\bibitem{5} P.Bambade, A.Dobrovolskaya, V.Novikov, Phys. Lett. {\bf
B319}, 348 (1993).
\bibitem{6} M.Voloshin, Phys. Rev. {\bf D47}, 2357 (1993); {\bf D47},
2573 (1993).
\bibitem{7} A. De Rujula et al., Nucl. Phys. {\bf B384}, 3 (1992).
\bibitem{8} A.Dobrovolskaya, V.Novikov, Z. Phys. {\bf Z57}, 865
(1993).
\end{thebibliography}
\end{document}